# Giant semiclassical magnetoresistance in high mobility TaAs$_2$ semimetal

Desheng Wu[1,2,†], Jian Liao[1,†], Wei Yi[1], Xia Wang[1], Peigang Li[2], Hongming Weng[1,*], Youguo Shi[1,*], Yongqing Li[1], Jianlin Luo[1,3], Xi Dai[1], Zhong Fang[1]

[1]Beijing National Laboratory for Condensed Matter Physics, Institute of Physics, Chinese Academy of Sciences, Beijing 100190, China

[2]Department of physics, Center for Optoelectronics Materials and Devices, Zhejiang Sci-Tech University, Hangzhou, 310018, China

[3]Collaborative Innovation Center of Quantum Matter, Beijing, China

## Abstract

We report the observation of colossal positive magnetoresistance (MR) in single crystalline, high mobility TaAs$_2$ semimetal. The excellent fit of MR by a single quadratic function of the magnetic field $B$ over a wide temperature range ($T$ = 2-300 K) suggests the semiclassical nature of the MR. The measurements of Hall effect and Shubnikov-de Haas oscillations, as well as band structure calculations suggest that the giant MR originates from the nearly perfectly compensated electrons and holes in TaAs$_2$. The quadratic MR can even exceed 1,200,000% at $B$ = 9 T and $T$ = 2 K, which is one of the largest values among those of all known semi-metallic compounds including the very recently discovered WTe$_2$ and NbSb$_2$. The giant positive magnetoresistance in TaAs$_2$, which not only has a fundamentally different origin from the negative colossal MR observed in magnetic systems, but also provides a nice complemental system that will be beneficial for applications in magnetoelectronic devices

Magnetoresistance (MR) is the relative change of longitudinal resistance of a material subjected to a magnetic field. MR has been observed in nearly all conducting materials and led to many important applications in devices such as magnetic sensors [1]. On the fundamental research side, MR can serve as a useful probe of the electronic structures of conductors including even superconductors for many decades [2-6]. In conventional nonmagnetic metals with closed Fermi surfaces, the MR tends to have a quadratic dependence on the low magnetic fields and saturates to a moderate value in the high field limit [7]. Such behavior can be understood within the semiclassical model whereby the large non-saturating MR, which is of great technological relevance, whilst can only be realized when the electrons are perfectly (or nearly perfectly) compensated by holes. Bismuth is the first such example in which the MR shows nonsaturating behavior even in high magnetic fields. However, behavior of the MR in bismuth shows clear deviation from the quadratic magnetic field dependence even when the magnetic field is only of a few Tesla [8, 9]. It was suggested as a consequence of low carrier density and small effective mass of the electrons [10,11], under which circumstance the Landau quantization in sufficiently strong magnetic field can drive the electron transport into a state of quantum limit, i.e., all electrons reside in the lowest Landau level. Recently, Ali *et al*. discovered that in a layered compound, $WTe_2$, the very large and quadratic MR could persist up to a magnetic field of $B = 60$ T and the MR ratio, defined as MR= $[\rho_{xx}(B) - \rho_{xx}(0T)]/ \rho_{xx}(0T) \times 100\%$, reaches $1.3 \times 10^5$ % at $B = 9$ T and $4.3 \times 10^6$ % at 60 T at 4 K [12]. The observation demonstrates the feasibility of searching for giant and non-saturating MR in semi-metallic compounds, alternative to the elemental semimetals such as bismuth, antimony and graphite [8, 9, 13, 14]. Further progress along this line may provide excellent candidate materials for magnetic devices that can work in both weak and strong magnetic fields, as well as for study on novel physics related to high carriers mobility, electron correlation and strong spin-orbit coupling [15].

In this work, the MR in our synthesized high quality $TaAs_2$ crystals is found to have quadratic dependence on the magnetic field and exceed $1.2 \times 10^6$ % at $B = 9$ T and $T = 2$ K. The remarkably high value is almost one order of magnitude larger than that of $WTe_2$. Both the MR and Hall resistance can be nicely fitted by the semiclassical two-band model for electron transport, suggesting that the large and quadratic MR arises from the nearly perfectly compensated electrons and holes with carrier densities close to 2.8 x $10^{18}$ cm$^{-3}$ and mobilities of about 1.2 x $10^5$ cm$^2$/V·s. First principles band structure calculations confirmed that TaAs$_2$ is indeed a semimetal in which a hole pocket and several electron pockets coexist. Analysis of Shubnikov de-Haas oscillations provided further evidences for the coexistence of multiple Fermi surfaces as well as their anisotropy.

Single crystals of TaAs$_2$ were grown by a chemical vapor transport method using I$_2$ as the carrier agent. The starting materials, Ta (99.99%), and As (99.999%), were put separately in two Al$_2$O$_3$ capsules which were sealed with appropriate amount of dry I$_2$ into a silica tube. The whole assembly was then slowly heated up to 1173 K and maintained at this temperature for seven days. It was subsequently cooled down to room temperature in 12 hours and many single crystals with different sizes were thus left in the capsule. The crystals were characterized by X-ray diffraction (XRD) on a Bruker SMART APEX II diffractometer using Cu $K_{\alpha 1}$ radiation ($\lambda$ = 1.5406 Å). Rietveld analysis was performed with powder XRD data collected on fine powder sample ground by using a TaAs$_2$ crystal. The program RIETAN-2000 was used for the analysis [16]. Magnetotransport measurements were performed with four-probe method in a Quantum Design PPMS and also in a helium vapor flow cryostat with temperatures down to 1.6 K and magnetic fields up to 9 T. Both dc and ac lock-in methods were used in the measurements.

The powder XRD pattern and the Rietveld analysis profiles are presented in Fig. 1 by the main panel. The analysis suggests a monoclinic structure (space group $C$2/m, no. 12) with $a$ = 9.3290 (12) Å, $b$ = 3.3846 (5) Å, $c$ = 7.7525 (7) Å and $\beta$ = 119.70 (2)°, which are close to the previously reported values [17, 18]. The typical size of TaAs$_2$ crystals, shown by an optical image in the upper inset, is about 1.5 × 1.0 × 0.6 mm$^3$. The top and an adjacent facets depicted in the upper-right inset, can be indexed as (001) and (-201) planes, respectively, based on the XRD patterns. The ohmic contacts were placed onto the large (001) surface (i.e. the top surface), and the current was applied along the [010] direction (the $b$ axis). The applied magnetic field was always perpendicular to the $b$ axis, and was also perpendicular to the top surface except the case of study on the transport anisotropy for which it was tilted out of the direction normal to the (001) plane.

Results of the transport measurements on TaAs$_2$ crystals are shown in Fig. 2 and the reliability of the data was confirmed by the similar results obtained from measurement on two independent crystals. When $B = 0$ T, the longitudinal resistivity $\rho_{xx}$ decreases rapidly as the temperature decreases from 300 K to about 20 K (Fig. 2a). The residual resistivity ratio of TaAs$_2$, $\rho_{300K}/\rho_{2K}$, is nearly 1000, comparable with the value reported for bismuth crystal which has been intensively studied aiming at discovery of various quantum and classical effects due to its high carriers mobility, small effective mass and long mean free paths of the electrons [19]. Seen in Fig.2a, it is apparent that $\rho_{xx}$ can be remarkably enhanced with $B$, especially at low temperatures. The low temperature $\rho_{xx}$ already shows a clear upturn at a small $B$ no larger than 0.3 T and is markedly enhanced with the further increase of $B$. As a contrast, the high temperature $\rho_{xx}$ changes gently and still behaves as a metal even when $B$ is increased up to 9 T. The behavior of $\rho_{xx}$ highly resembles that observed previously in both bismuth and graphite in which the mechanism was yet debated between the model of multiple types of carriers in the semi-metallic systems [19] and metal-insulator transition induced by magnetic field [20, 21].

The linear behavior of log(MR) versus log$B$ in Fig. 2b indicates that the MR has a quadratic dependence on $B$ over the measured temperature range. The MR at $B = 9$ T and T= 300 K is only 12%, but it increases significantly to 1,200,000% at 2 K when $B$ is the same. The well-fitting curve to the $B$ dependent MR in Fig. 2c by using a single quadratic function with the expression MR($B$)/MR(9 T) = $\alpha B^2$ at various temperatures further demonstrates the quadratic dependence of MR on $B$. The fact that the MR follows the Kohler's rule over such a wide temperature range strongly suggests that it has a semiclassical origin, similar to that of bismuth single crystal in low magnetic fields.

In the semiclassical two-band model [22], the longitudinal resistivity $\rho_{xx}$ and the Hall resistivity $\rho_{xy}$ are given by

$$\rho_{xx} = \frac{\sigma_e + \sigma_h + \sigma_e \sigma_h (\sigma_e R_e^2 + \sigma_h R_h^2) B^2}{(\sigma_e + \sigma_h)^2 + \sigma_e^2 \sigma_h^2 (R_e + R_h)^2 B^2}$$

$$\rho_{xy} = B \cdot \frac{R_e \sigma_e^2 + R_h \sigma_h^2 + \sigma_e^2 \sigma_h^2 R_e R_h (R_e + R_h) B^2}{(\sigma_e + \sigma_h)^2 + \sigma_e^2 \sigma_h^2 (R_e + R_h)^2 B^2}$$

where $\sigma_e = n_e e \mu_e$, $R_e = -1/n_e e$, $n_e$, and $\mu_e$ ($\sigma_h = n_h e \mu_h$, $R_h = -1/n_h e$, $n_h$ and $\mu_h$) are longitudinal conductivity, Hall coefficient, carrier density, and mobility for electrons (holes), respectively. In case of perfect compensation between electrons and holes, i.e. $R_e+R_h = 0$, MR has a simple quadratic dependence on $B$ in a formula MR = $\mu_e\mu_h B^2$. Slight deviation from the perfect resonant condition can also lead to large quadratic MR as long as $B$ is not too strong. The mechanism can reasonably explain the data depicted in Fig. 2b-c. Fig. 2d and 2e further demonstrate that both $\rho_{xx}$ and $\rho_{xy}$ can be nicely fitted by the semiclassical two-band model.

The extracted electron and hole densities from the fits are $n_e$ = 2.790±0.001 x $10^{18}$ cm$^{-3}$ and $n_h$ = 2.786±0.001 x $10^{18}$ cm$^{-3}$, respectively. The small difference (less than 0.2%) between the two values implies that the two types of carries are almost fully compensated with each other. The calculated carrier mobilities are $\mu_e$ = 1.207×10$^5$ cm$^2$/V·s and $\mu_h$ = 1.208 × 10$^5$ cm$^2$/V·s, respectively, indicating that TaAs$_2$ has remarkably high mobilities. The high mobilities and nearly perfect carriers compensation in TaAs$_2$ should be the major sources for the unprecedentedly large and quadratic MR.

The preliminary first-principles band structure calculations on the electronic structure of TaAs$_2$ using the OpenMX [26] software package can explain the semimetallic nature. The choice of pseudo atomic orbital basis set with Ta9.0-$s2p2d2f1$ and As9.0-$s2p2d1$, the pseudo-potential and the sampling of Brillouinzone (12 ×12 ×6 k-grid) have been checked. The exchange-correlation functional within generalized gradient approximation parameterized by Perdew, Burke, and Ernzerhof [27] has been used. Spin-oribt coupling is included self-consistently. As depicted in Fig. 3, there are several overlaps between the conduction and the valence bands. As a result, one hole pocket surrounding the L (0.5, 0.5, 0.5) point coexists with a few pairs of electron pockets near the L point and the Y (0,0.5,0) point. It is interesting to note that both the highest occupied valence band and the lowest unoccupied conduction band are mostly composed of Ta 5d orbitals. These 5d bands are quite wide and we have confirmed that the nonmagnetic state is stable. In case of perfect stoichiometry, the carrier density of electrons would be exactly the same as the hole density. The existence of multiple electron pockets with different sizes, however, makes the quantitative comparison with the experiment not straightforward. This is further complicated by the anisotropic Fermi surfaces of both electrons and holes. All of them are of elliptical shape, as shown in Fig. 3b-d.

The feature of multiple Fermi surfaces of TaAs$_2$ is verified by the Shubnikov-de Haas (SdH)

oscillations that superimpose on the semiclassical MR. Fig. 4a shows the SdH signals in magnetic fields up to $B$ = 9T after subtraction of the four-order polynomial background . The beating pattern implies multiple types of carriers of high mobility. This can be seen more clearly by the fast Fourier transform (FFT) analysis shown in Fig. 4b in which the most prominent pairs of peaks correspond to frequencies of $f_1$ = 190.5 T and $f_2$ = 212.5 T. The second and third harmonics of these two peaks can also be observed in the same figure. In addition, much smaller peaks can also be displayed at frequencies lower than $f_1$ and $f_2$ but larger than $\Delta f = f_2 - f_1$, which suggests that these small peaks may originate from the Fermi pockets much smaller than those corresponding to $f_1$ and $f_2$. This is qualitatively in agreement with the band structure calculation results shown in Fig. 3. From the temperature dependence of the SdH oscillations amplitude, the effective mass of 0.34$m_0$ and 0.30$m_0$ can be obtained for 190.5T and 212.5T Fermi pockets, respectively, where $m_0$ is the free electron mass. It should be, however, taken as a rough estimation of the effective mass of the electrons in this complicated material. The Fermi surfaces are expected to be anisotropic from the band structure calculations. It can also be confirmed by the experimental results by the differences in the SdH oscillations curves shown in Fig. 4c which is taken in magnetic fields with different tilting angles. The electronic anisotropy is more clearly displayed by the corresponding FFT patterns (Fig. 4d) which vary clearly with the tilting angles. Detailed analysis of the Fermi surfaces would require extensive measurements of the MR in high magnetic fields systematically tilted in azimuth angles because of the complicated crystal structure of TaAs$_2$. This is out of scope of the present work and will be addressed elsewhere.

From the above measurements, it is clear that a prominent feature of TaAs$_2$ is the high carriers mobilities of $10^5$ cm$^2$/V·s which is an order of magnitude higher than the largest values of bulk GaAs [23], even though the effective mass of electrons of n-GaAs is only 1/6 of that in TaAs$_2$. This fact suggests very weak impurity scattering of the carriers in TaAs$_2$, which can be attributed to the possible very low defect and impurity densities, as well as screening effects of electrons and holes. It is well known that, in bulk GaAs, a minimum electron density is needed to overcome the localization effects [24] but at a cost of introducing some amount of charged impurities, which in turn limits the carrier mobility. In contrast, the finite densities of electrons and holes created by band overlapping in semimetals would not disappear even when the samples are free of impurities or defects. Therefore, improvement of the crystal quality may further increase the carrier mobility and the level of carrier compensation in $TaAs_2$. This in principle can lead to very large quadratic MR that can survive to very high magnetic fields. Nevertheless, the low temperature MR in $TaAs_2$ achieved in this work has already reached $1.2 \times 10^6$ % at $B = 9$ T, approximately one order magnitude higher than the value obtained at the same magnetic field in the recently discovered $WTe_2$ [12] and $NbSb_2$ [25]. Despite the many similarities in the observed MR of $TaAs_2$ as those of $WTe_2$ and $NbSb_2$, quantitative fits to the semiclassical two-band model have never been reported for the latter two compounds [12, 25]. Nevertheless, no consensus has been reached on the mechanisms for the giant MR in $WTe_2$ and $NbSb_2$. The giant MR in $WTe_2$ was attributed to the semiclassical effects involving carrier compensation [12], whilst that in $NbSb_2$ was suggested to be a quantum mechanical effect related to the Dirac-like point [25]. The well-fitting curve of the giant MR in $TaAs_2$ to the semiclassical two-band model is consistent with the model for $WTe_2$. It is noteworthy that the spin-orbit coupling and the related spin texture may affect the giant MR, as pointed out in $WTe_2$ [28]. Further investigation is needed to clarify the influence. The results are valuable for providing important insights into the mechanisms underlying the giant MR in semimetals. Additionally, the very large positive MR, which is comparable with the large negative MR in colossal MR materials in magnitude, is of utmost importance for technical use.

**Acknowledgments**

This research was supported in part by the National Natural Science Foundation of China (No. 11274367, 11474330, 61274017), and the Strategic Priority Research Program (B) of the Chinese Academy of Sciences (No. XDB07020000), and the "International Young Scientist Fellowship (No. 2014004)" of IOP, CAS.

†These two authors contributed equally to this work.
*ygshi@iphy.ac.cn, hmweng@iphy.ac.cn

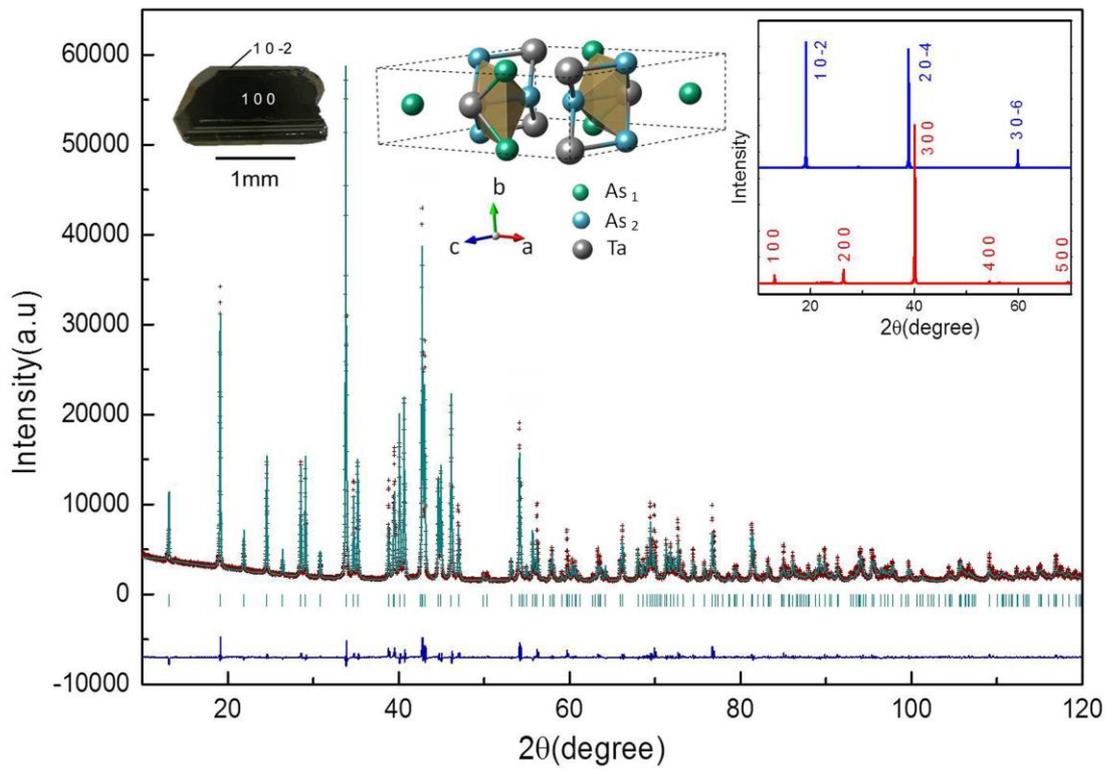

Figure 1. The experimental XRD pattern of TaAs$_2$ and the Rietveld analysis profiles. Cross markers and solid lines show the observed and calculated XRD patterns, respectively, and the difference is shown at the bottom. The upper-left insets an optical image of a TaAs$_2$ single crystal (sample #1), on which two surfaces [(001) and (-201)] are indexed with the XRD measurements shown by the upper-right inset. The upper-middle inset depicts the unit cell of TaAs$_2$.

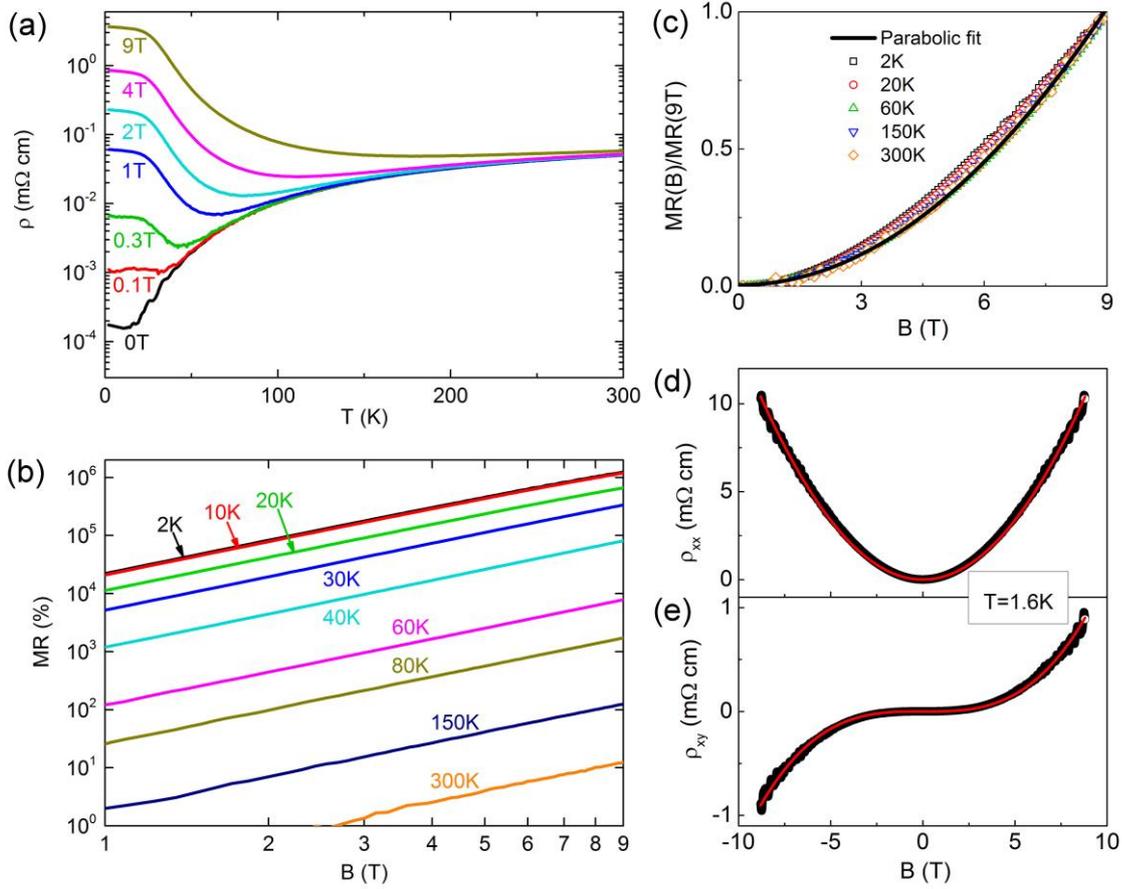

Figure 2. (a) Temperature dependence of the longitudinal resistivity $\rho_{xx}$ in various magnetic fields ($B$ = 0-9 T). (b) The MR plotted as a function of magnetic field at temperatures from 2 to 300 K. (c) Normalized MR, i.e. MR($B$)/MR (9T), at various temperatures (the symbols). All of the MR curves can be fitted by a single quadratic function (the solid line gives such a fit of the MR at 300K and there is a little difference in the line shape of the fitting curves). (d) The longitudinal resistivity $\rho_{xx}$ and (e) the Hall resistivity $\rho_{xy}$ at $T$= 1.6 K and the fitted results by the semiclassical two-band model (the symbols are the experimental data and the solid line is the fitted result). The sign of $\rho_{xy}$ for electrons is set to be positive. The data in (a) - (c) and (d) - (e) were taken from samples #1 and #2, respectively.

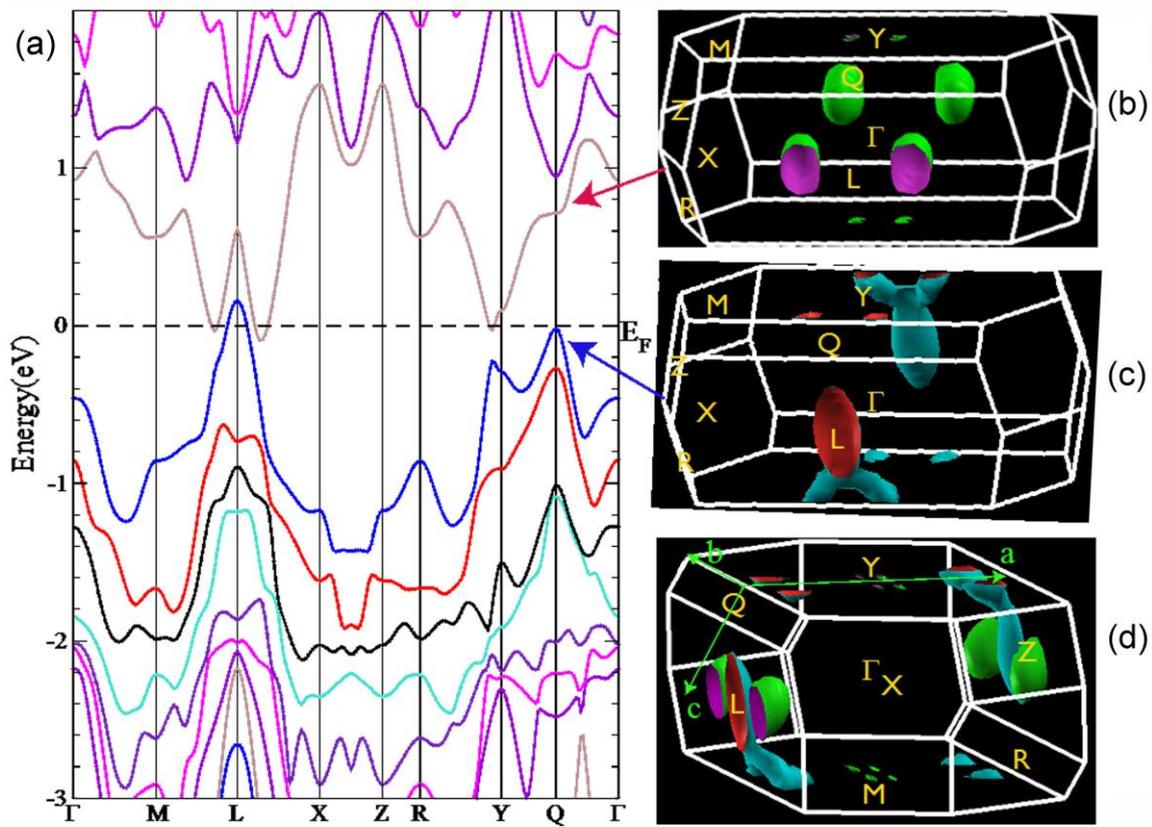

Figure 3. (a) Band structure of TaAs$_2$ calculated by GGA with including spin-orbit coupling. (b) The Fermi surface for the lowest partially occupied conduction band. (c) The Fermi surface for the highest partially occupied valence band. (d) The merged Fermi surfaces in (b) and (c). The high symmetrical k-point is labeled in the Brillouin zone (BZ). The planes in BZ parallel to the lattice constants *a*, *b* and *c* in marked with arrows.

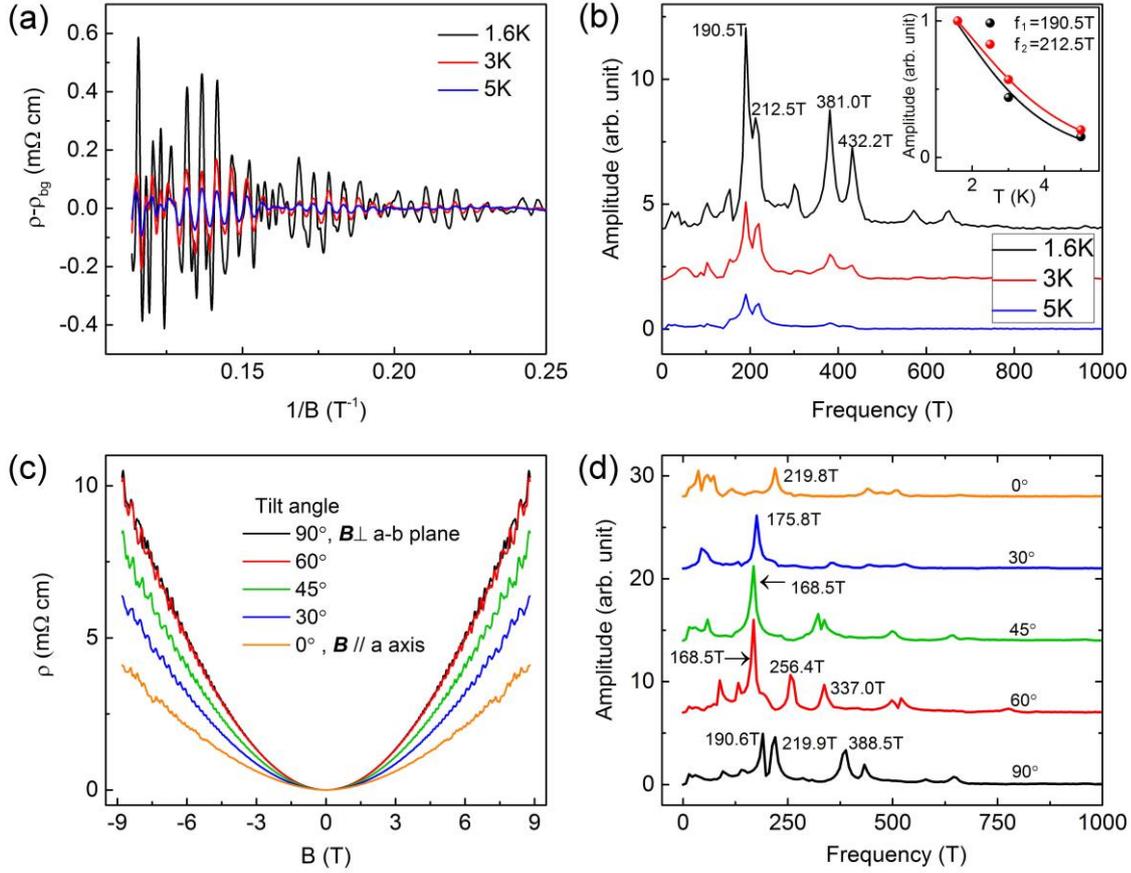

Figure 4. (a) SdH oscillations curves obtained by subtracting a four-order polynomial background at 1.6 K, 3 K, and 5 K. (b) FFT analysis of the SdH oscillations shown in the main panel. The upper inset shows extraction of the effective mass of the electrons from the temperature dependence of SdH amplitude with Lifshitz-Kosevich formula. (c) Resistivity vs. magnetic field when magnetic fields have various tilting angles at $T$ = 1.6 K. The magnetic field was applied in the ac-plane and the tilting angle $\theta$ is the angle of magnetic field relative to the $a$ axis (i.e. [100] direction). (d) FFT spectra of the quantum oscillations for various magnetic field tilts shown in panel c. The curves are vertically shifted for clarity. All of the data were taken from sample #2.